\documentclass[pdflatex,sn-basic]{sn-jnl}

\usepackage{graphicx}%
\usepackage{multirow}%
\usepackage{amsmath,amssymb,amsfonts}%
\usepackage{amsthm}%
\usepackage{mathrsfs}%
\usepackage[title]{appendix}%
\usepackage{xcolor}%
\usepackage{textcomp}%
\usepackage{manyfoot}%
\usepackage{booktabs}%
\usepackage{algorithm}%
\usepackage{algorithmicx}%
\usepackage{algpseudocode}%
\usepackage{listings}%
\usepackage{subcaption}

\graphicspath{{figures/}}

\newcommand{\muares}{\textmu Ares\ }

\newcommand{\msun}{\mathrm{M_{\odot}}}

\theoremstyle{thmstyleone}%
\theoremstyle{thmstyletwo}%

\theoremstyle{thmstylethree}%

\begin{document}

\title[Multiband MBHBs]{Archival Multiband Gravitational-Wave Signals from Massive Black Hole Binary Mergers}


\author*[1,2]{\fnm{Alexander~W.} \sur{Criswell}}\email{alexander.criswell@vanderbilt.edu}

\author[1]{\fnm{Stephen~R.} \sur{Taylor}}\email{stephen.r.taylor@vanderbilt.edu}

\author[3]{\fnm{Kris} \sur{Pardo}}\email{kmpardo@usc.edu}
\author[4,5,6]{\fnm{Alberto} \sur{Sesana}}\email{alberto.sesana@unimib.it}
\author[7]{\fnm{David} \sur{Izquierdo}}\email{dizquierdo@ice.csic.es}
\author[8,9]{\fnm{Silvia} \sur{Bonoli}}\email{silvia.bonoli@dipc.org}
\author[10]{\fnm{Daniele} \sur{Spinoso}}\email{dspinoso@.outlook.it}

\affil[1]{\orgdiv{Department of Physics \& Astronomy}, \orgname{Vanderbilt University}, \orgaddress{\street{ 2201 West End Ave}, \city{Nashville}, \postcode{37235}, \state{TN}, \country{USA}}}

\affil[2]{\orgdiv{Department of Life and Physical Sciences}, \orgname{Fisk University}, \orgaddress{\street{1000 17th Avenue N.}, \city{Nashville}, \postcode{37208}, \state{TN}, \country{USA}}}

\affil[3]{\orgdiv{Department of Physics \& Astronomy}, \orgname{University of Southern California}, \orgaddress{\street{825 Bloom Walk}, \city{Los Angeles}, \postcode{90089}, \state{CA}, \country{USA}}}

\affil[4]{\orgdiv{Dipartimento di Fisica ``G. Occhialini"}, \orgname{Universit{\'a} degli Studi di Milano-Bicocca}, \orgaddress{\street{Piazza della Scienza 3}, \city{Milano}, \postcode{20126}, \country{Italy}}}

\affil[5]{\orgdiv{INAF}, \orgname{Osservatorio Astronomico di Brera}, \orgaddress{\street{via Brera 20}, \city{Milano}, \postcode{20121}, \country{Italy}}}

\affil[6]{\orgdiv{INFN}, \orgname{Sezione di Milano-Bicocca,}, \orgaddress{\street{Piazza della Scienza 3}, \city{Milano}, \postcode{20126}, \country{Italy}}}

\affil[7]{\orgdiv{Institute of Space Sciences (ICE, CSIC)}, \orgname{ICE-CSIC}, \orgaddress{\street{Campus UAB, Carrer de Magrans}, \city{Barcelona}, \postcode{08193}, \country{Spain}}}

\affil[8]{ \orgname{Donostia International Physics Centre (DIPC)}, \orgaddress{\street{Paseo Manuel de Lardizabal 4}, \city{Donostia-San Sebastian}, \postcode{20018}, \country{Spain}}}

\affil[9]{ \orgname{IKERBASQUE, Basque Foundation for Science}, \orgaddress{\city{Bilbao}, \postcode{E-48013}, \country{Spain}}}

\affil[10]{\orgdiv{Como Lake Center for Astrophysics}, \orgname{Universit{\'a} degli Studi dell'Insubria}, \orgaddress{\street{Via Valleggio 11}, \city{Como}, \postcode{22100}, \country{Italy}}}


\abstract{While massive black hole binaries (MBHBs) merge at gravitational-wave frequencies above the pulsar timing array (PTA) sensitivity band, we show that they leave orphaned low-frequency contributions in the PTA pulsar term. Due to the light-propagation time between each pulsar in the array and Earth, the pulsar term acts as a time-delayed probe of a chirping merger with a specific frequency response determined by the direction of origin and intrinsic properties of the MBHB. We provide a detailed consideration of how such a multiband signal would manifest in a full PTA, demonstrate an approach to stack these orphaned pulsar terms across the array, and discuss prospects for an archival, multiband search in conjunction with MBHB mergers observed in astrometric data or spaceborne interferometers like LISA.}




\maketitle

\section{Introduction}\label{sec:intro}
The hierarchical assembly of galaxies \citep{WhiteandRees1978,WhiteFrenk1991}, together with the near-ubiquitous presence of massive black holes (MBHs) at their centres \citep[see e.g,][]{miller_xray_2015}, implies that massive black hole binaries (MBHBs) should be a natural outcome of galaxy mergers \citep{Begelman1980}. Under a theoretical perspective, it has been shown that after galaxy interactions, the two central MBHs gradually migrate toward the center of the remnant galaxy via dynamical friction, eventually forming a gravitationally bound system \citep{Chandrasekhar1943}. The subsequent evolution of the binary is governed by interactions with the surrounding stellar population, any circumbinary gas disc, as well as by the emission of gravitational waves, which ultimately drives the binary to coalescence \citep{Quinlan1997,Sesana2006,Escala2005,Dotti2007,Cuadra2009,Franchini2021}. 

Despite decades of theoretical work, MBHBs have remained elusive in the electromagnetic spectrum, as their parsec to sub-parsec separations and long orbital periods hinder any direct imaging and prevent unambiguous detection. To date, several candidate MBHBs have been identified through periodic or quasi-periodic variability in active galactic nucleus (AGN) light curves, potentially reflecting orbital motion or accretion modulations induced by the binary \citep{GrahamM2015,DOrazio2018,DeRosa2019,DOrazio2023, Luo2025, TubinArenas2025,rodriguez_compact_2006}. However, discriminating genuine MBHB signatures from AGN variability remains a major challenge as it demands long-term, high-cadence monitoring combined with rigorous statistical analysis \citep[see e.g,][]{Bertassi2025,Mohammed2025}.

Low-frequency gravitational wave observations provide an extremely promising route forward to expand our understanding of MBHB astrophysics. In the nHz, pulsar timing arrays (PTAs) have detected strong evidence for a gravitational-wave background \citep{NANOGrav:2023gor, antoniadis_second_2023b, xu_searching_2023, zic_parkes_2023, miles_meerkat_2025}. A likely source of this background is the population of supermassive black hole binaries (SMBHBs), thereby providing a unique insight into the high-mass end of the MBHB spectrum and the galaxies which host them. Indeed, analyses of the nHz gravitational-wave background have begun to constrain the astrophysics of massive binary formation and large galaxy mergers \citep{agazie_nanograv_2023n, harris_connecting_2024,2024arXiv241105906C,bi_implications_2023a, ellis_gravitational_2024, liepold_big_2024, sato-polito_where_2024}. 

Looking forward, one of the most exciting prospects for PTA science is detection of individual SMBHBs, which would manifest as monochromatic continuous wave signals in the PTA data \citep{2010PhRvD..81j4008S,2009MNRAS.394.2255S,2013CQGra..30v4004E,2012ApJ...756..175E,2014PhRvD..90j4028T,2016ApJ...817...70T,2010arXiv1008.1782C}. Searches for individual SMBHBs in PTA datasets have taken place, but only upper limits have been placed \citep{agazie_nanograv_2023o,agarwal_nanograv_2026,antoniadis_second_2024, falxa_searching_2023, miles_meerkat_2025a, zhao_searching_2025}. Detection of such a signal would not only confirm the existence of close (sub-pc) SMBHBs and their necessary contribution to the nHz gravitational-wave background, but also provide opportunity for detailed multimessenger characterization of the SMBHB and its host galaxy \citep[e.g.,][]{2024ApJ...976..129P,2025arXiv251215911B,2025arXiv251008683C,2025ApJ...990...46V,2025CQGra..42b5021C,2024PhRvL.132f1401C,2021ApJ...921..178L,2023ApJ...945...78L,2026A&A...706A.115T}. The upcoming 3$^{\rm rd}$ data release of the International Pulsar Timing Array (IPTA) \citep{2023AAS...24143802G} will be the most sensitive PTA dataset to-date and has the potential to provide new insights into the dynamics and demographics of SMBHBs \citep[e.g.,][]{2017MNRAS.468..404C,2019MNRAS.488..401C,2017PhRvL.118r1102T,2025arXiv251211981L,2024A&A...687A..42B,2025ApJ...982...55L}. 

Whereas PTAs have access to a gravitational-wave background and can, in the future, probe individual, slowly evolving systems, spaceborne interferometers have access to these binaries' final inspiral and merger. The Laser Interferometer Space Antenna (LISA), a joint ESA-NASA mission with a scheduled launch in 2035, is expected to observe MBHB mergers with total masses ranging from $10^4-10^8\,\msun$ \citep{colpi_lisa_2024}. LISA will greatly expand our understanding of these systems and their host galaxies, illuminating their origin, growth, and evolution over cosmic time \citep{langen_hierarchical_2025a, mangiagli_massive_2025, sesana_reconstructing_2011}; see \citet{amaro-seoane_astrophysics_2023e} and references therein for a review.

Between the supermassive black hole binaries in the nHz band and the somewhat lighter systems in the mHz lies the domain of both proposed and ongoing experiments. In the \textmu Hz, several concepts have been proposed in response to the European Space Agency's Voyage 2050 call for the next generation of large missions \citep{sesana_unveiling_2021,martens_lisamax_2023}, including \muares \citep{sesana_unveiling_2021}. A LISA successor spanning the Martian orbit, \muares would have the unprecedented ability to observe and characterize high-mass MBHB mergers via deep sensitivity in the \textmu Hz band.

In the present day, ongoing experiments amenable to the astrometric detection of gravitational waves provide a possible avenue to detect \textmu Hz sources. As with PTAs, astrometric gravitational wave detection relies on the correlated change to photon trajectories induced by a gravitational wave at Earth \citep{Book2011}. The transverse distortions to the photon trajectories result in small changes to the apparent positions of point sources on the sky, which can be measured using astrometric surveys, such as \textit{Gaia} \citep{Klioner2018, Moore2017, Darling2025}. However, detecting individual MBHB mergers would require higher frequencies than current \textit{Gaia} data could probe. Recent work has suggested the use of relative astrometric detection to push to higher frequencies \citep{Wang2021, Wang2022, Pardo2023}. This method measures changes in star positions relative to their own stellar mean position over the time of a survey, which allows for the use of any photometric survey with good
photometric stability, angular resolution, and cadence. Two promising telescopes with the requisite properties to possibly observe MBHB mergers \citep{Zhang2025} are the \textit{Kepler} Space Telescope (``\textit{Kepler}'')\citep{Borucki2010} and the \textit{Nancy Grace Roman} Space Telescope (``\textit{Roman}'') \citep{gaudi_2019_AuxiliaryScienceWFIRST}.

The rich present and future landscape of low-frequency gravitational-wave astronomy provides a compelling opportunity for combining information from observations across several frequency bands: ``multiband" gravitational-wave astronomy. Indeed, the idea of multiband gravitational-wave observations is far from new; such approaches have been explored in the context of LISA and terrestrial observatories such as LIGO-Virgo-KAGRA \citep{aasi_advanced_2015, acernese_advanced_2015b, akutsu_kagra_2019}, proposed next-generation facilities Einstein Telescope \citep{punturo_einstein_2010} and Cosmic Explorer \citep{reitze_cosmic_2019a}. Following the discovery of the first stellar-mass binary black hole merger observed in gravitational waves, GW150914 \citep{abbott_observation_2016}, it was noted that LISA would have confidently detected the inspiraling system well before merger --- had it been observing at the time \citep{sesana_prospects_2016d}. A number of studies since have focused on the considerable scientific promise and potential challenge of such multiband gravitational-wave observations
\citep{buscicchio_stellarmass_2025, jani_detectability_2020, seto_how_2022b, toubiana_detectability_2022, wong_expanding_2018, romero-shaw_eccentric_2024,wu_multiband_2025}. In particular, several studies \citep{wong_expanding_2018,ewing_archival_2021, wang_spacebased_2024} have noted the specific potential for \textit{archival} multiband searches: using a detected merger in terrestrial observatories to perform a targeted search in LISA data and extract otherwise undetectable low signal-to-noise ratio (SNR) signals. Finally, inspired by \citet{pitkin_there_2008} --- which considers in brief a related concept to that of this work, focused on pulsar distance estimates --- \citet{spallicci_complementarity_2013} considered the low-frequency multiband case of sequential detection: a MBHB system first observed in PTAs, which would then evolve over the course of $\sim$decades and eventually merge in LISA. However, even under the most optimistic of assumptions the probability of such a system both existing and also happening to transit between the two frequency regimes during the observational window is exceedingly small.

\section{Orphaned Pulsar Terms}\label{sec:concept}
From the results of \citet{spallicci_complementarity_2013}, we can conclude that the only way to achieve a decent probability of multiband observations of a MBHB system is to allow for hundreds or even thousands of years to pass between PTA observations and observation of the higher-frequency merger. This is, of course, impractical. However, the very nature of PTA gravitational-wave detection provides the necessary mechanism.

Formally, the gravitational-wave signature in a PTA can be divided into two components: the Earth term and the pulsar term. The Earth term describes deflections of the position of Earth with respect to the pulsars in the array, whereas the pulsar term describes the influence of propagating gravitational waves on each individual pulsar. While the majority of searches do account for pulsar terms in the signal model \citep[e.g.,][]{2010arXiv1008.1782C,2013CQGra..30v4004E,2025PhRvD.112h3035G,2022PhRvD.105l2003B}, current-day PTA searches recognize that the Earth term will likely be the primary source of information for the first SMBHB detections \citep{2024PhRvL.132f1401C}. Including the pulsar term requires knowledge or parameterization of the distance to each pulsar in the array, and pulsar distances are currently poorly constrained in most cases \citep[e.g.,][]{manchester_australia_2005}. As PTA data becomes more informative, neglecting the pulsar term could introduce sky location and other parameter biases, hence it is modeled with pulsar distances marginalized over as nuisance parameters with priors set at current measurement precisions. In the future, PTA data itself could refine these pulsar distances, albeit with constraints that are covariant with the detected system's binary parameters \citep{2010arXiv1008.1782C,2013CQGra..30v4004E,2025arXiv251210795T,Yu:2025tnk,Kato:2025set,Xiao:2025mcg}.

If we consider, however, the case of a MBHB merger detected by a higher-frequency observatory within the Solar System, the scenario becomes quite different. The Earth term of such a signal is what is being detected at merger; it provides a high-frequency contribution well above the nHz and, as such, is negligible within the PTA datastream. However, the light received at Earth from a given pulsar contemporaneously with the merger does not trace the gravitational-wave perturbation at Earth. Instead, it carries information about the gravitational wave (and thus the binary dynamical state) from when it had passed by each pulsar. Pulsars in the array are at kpc distances; the pulsar term contribution therefore contains the gravitational wave signal as it existed approximately thousands of years prior to merger. For SMBHBs in the PTA band, accounting for the time-delay between the Earth term and pulsar terms allows for characterization of the system's evolution over time \citep{jenet_constraining_2004,sesana_measuring_2010,corbin_pulsar_2010,2012PhRvL.109h1104M}. For a chirping MBHB of sufficient mass, however, this time-delay is realized as a gravitational-wave signal entirely absent in the Earth term, but which lingers on at nHz frequencies within every pulsar in the array: an orphaned pulsar term.

\begin{figure}
    \centering
    \includegraphics[width=0.6\linewidth]{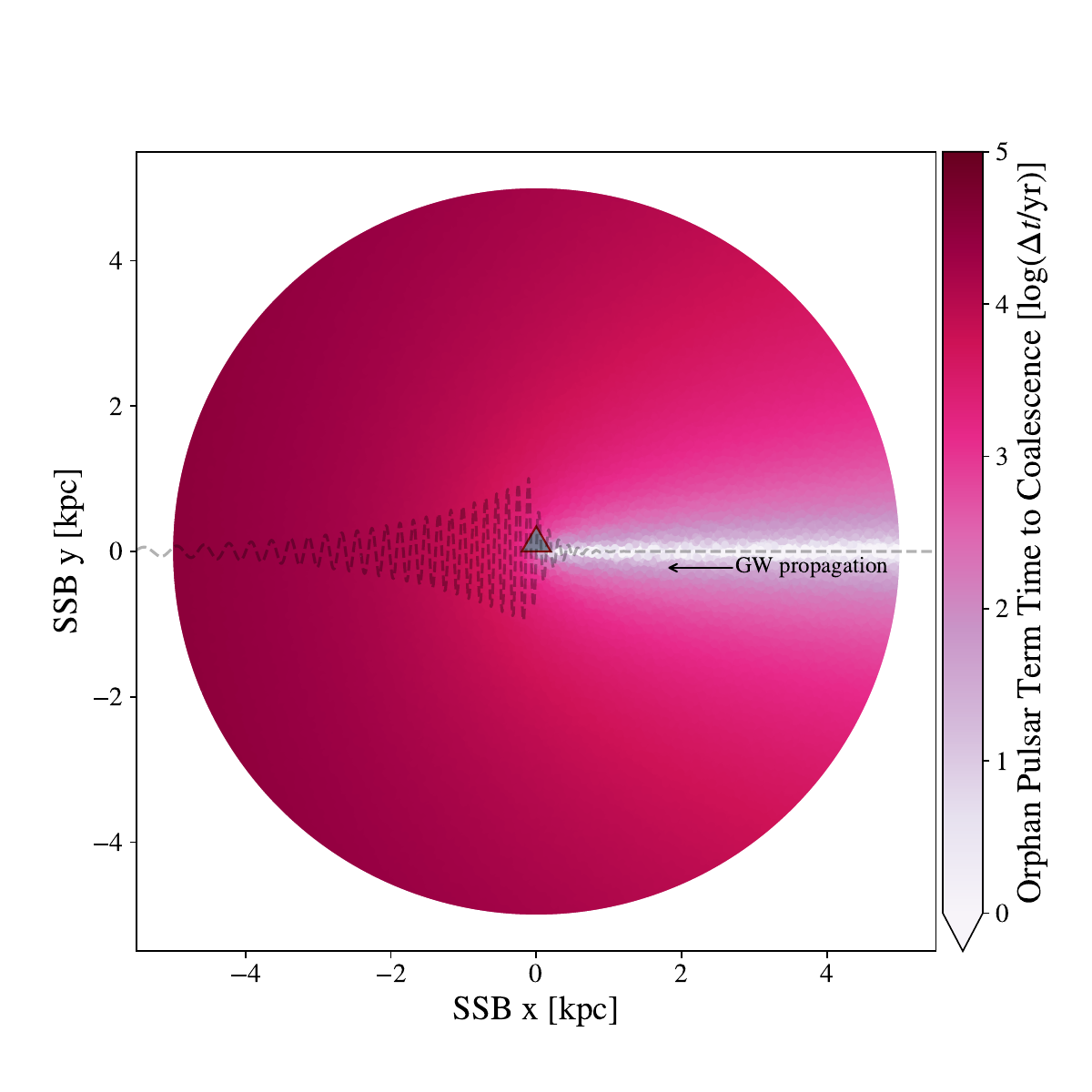}
    \caption{Time to coalescence of a MBHB system as seen in the pulsar term of pulsars in a hypothetical PTA as a function of position relative to Earth and the direction of incidence of the gravitational wave. At time of merger as observed at Earth, light arriving from pulsars antipodal to the direction of gravitational-wave origin carries information on the MBHB as it existed ten of thousands of years before merger. 
    }
    \label{fig:time-delay-map}
\end{figure}

Mathematically, we can understand the orphaned pulsar term mechanism as follows. The fractional shift to the arrival rate of pulses induced by a gravitational plane wave with direction of propagation $\hat{k}$ is
\begin{equation}\label{eq:pta_response}
    z(t) \equiv \sum_{A} \frac{1}{2}\frac{\hat{p}^a\hat{p}^b}{1+\hat{p}\cdot\hat{k}}\epsilon^{A}_{ab}(\hat{k},\psi) \left( h^{ab}_{A,\oplus} - h^{ab}_{A,\mathrm{psr}}e^{-i2\pi f_{\rm psr}D(1+\hat{k}\cdot\hat{p})/c}   \right),
\end{equation}
where $\hat{p}$ is a unit vector pointing from Earth to the pulsar, $f_{\rm psr}$ is the gravitational-wave frequency at the pulsar, $D$ is the distance to the pulsar, $\epsilon_{ab}^{A}$ are the polarization tensors for gravitational-wave polarizations $A\in\{+,\times\}$ given a polarization angle $\psi$, and $h^{ab}_{A,\oplus} = h^{ab}_{A,\oplus}(t,\phi_{0})$ and $h^{ab}_{A,\mathrm{psr}} = h^{ab}_{A,\mathrm{psr}}(t_{\rm psr},\phi_{0})$ are the gravitational wave contributions at Earth and the pulsar, respectively. The left-hand term in the parenthetical describes the Earth term, and on the right is the pulsar term. In the case of the orphaned pulsar terms, the Earth term is negligible, and we can write
\begin{equation}\label{eq:ophan_response}
    z(t) \equiv \sum_{A} \frac{1}{2}\frac{\hat{p}^a\hat{p}^b}{1+\hat{p}\cdot\hat{k}}\epsilon^{A}_{ab}(\hat{k},\psi) \left(- h^{ab}_{A,\mathrm{psr}}e^{-i2\pi f_{\rm psr}D(1+\hat{k}\cdot\hat{p})/c}   \right).
\end{equation}
From Eq.~\eqref{eq:ophan_response}, we can note two important effects for the orphaned pulsar terms. First, the time delay of the orphaned pulsar term with respect to the merging MBHB is dependent on both the pulsar distance and angle of incidence of the gravitational wave with respect to a given pulsar:
\begin{equation}\label{eq:orphan_time_delay}
    \Delta t_{\rm orphan} = (1+\hat{k}\cdot\hat{p})\frac{D}{c}.
\end{equation}
The time delay is therefore maximized for pulsars which are both far from Earth and antipodal to the origin of the gravitational wave. One can understand this effect by noting that the gravitational wave must first travel to a given pulsar, and then the light from that pulsar must travel back to Earth, yielding a factor of two for antipodal pulsars. The orphaned pulsar term time-delay as a function of both of these factors is depicted in Fig.~\ref{fig:time-delay-map}. Second, however, we must note that the overall PTA response carries a factor of $(1+\hat{p}\cdot\hat{k})^{-1}$, reducing its sensitivity in pulsars antipodal to the gravitational wave origin.

\begin{figure}
    \centering
    \includegraphics[width=0.6\linewidth]{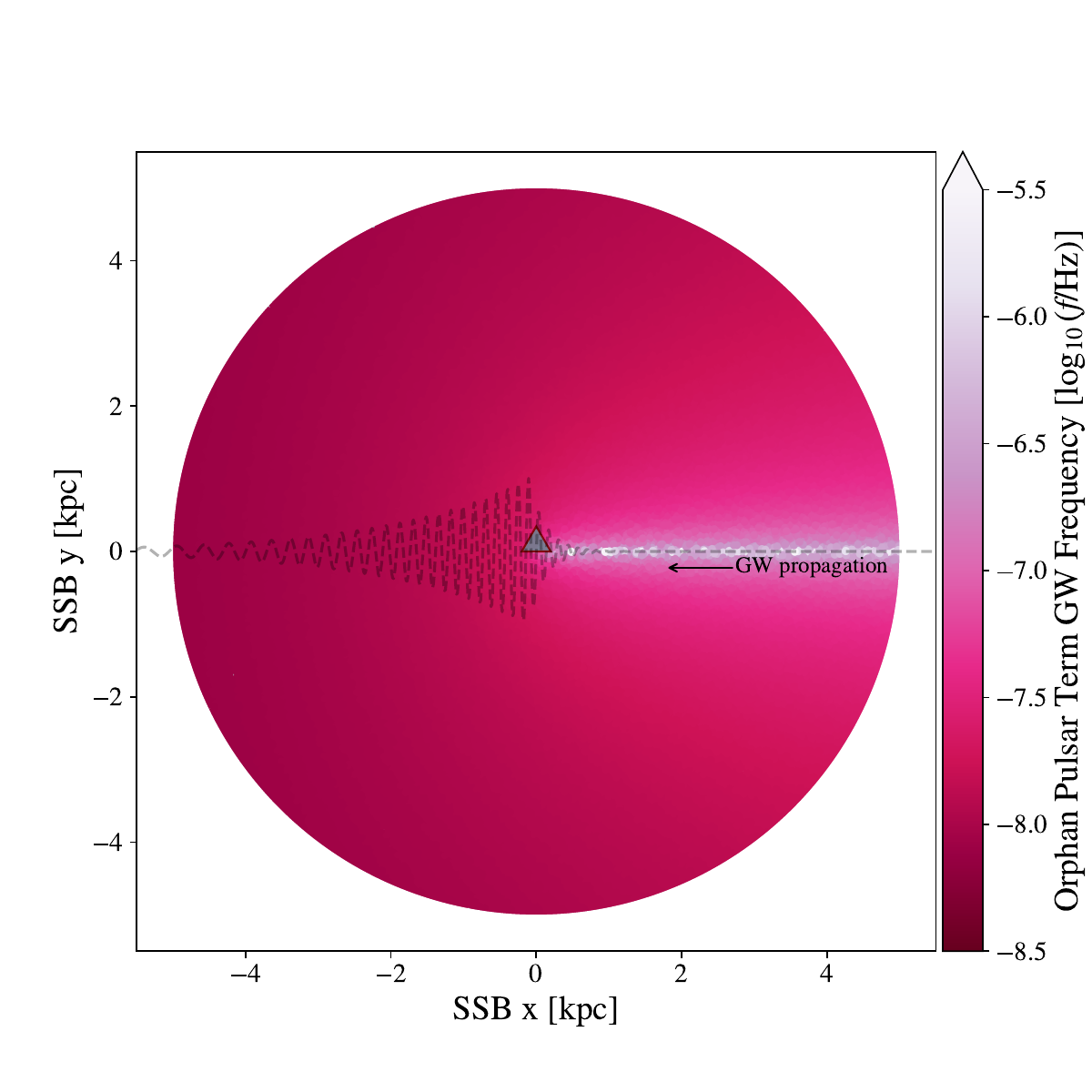}
    \caption{Frequency of the MBHB orphaned pulsar term for a system with total mass $\mathrm{M} = 2\times10^9\,\msun$ at redshift $z=1$, in a hypothetical PTA as a function of position relative to Earth and the direction of incidence of the gravitational wave. In distant pulsars well-separated on the sky from the direction of gravitational wave origin, the orphaned pulsar terms have gravitational-wave frequencies within the PTA band. 
    }
    \label{fig:frequency-shift-map}
\end{figure}

Crucially, the time delay of the orphaned pulsar terms corresponds to a frequency shift in the observed signal due to gravitational-wave-driven evolution of the system. For a circular MBHB with chirp mass $\mathcal{M}$ evolving under vacuum general relativity (GR), we can write the standard expression for the gravitational-wave frequency as a function of time to coalescence $t_c$:
\begin{equation}\label{eq:foft}
    f(t) = \frac{5^{3/8}}{8\pi}\left(\frac{G\mathcal{M}}{c^3}\right)^{-5/8}\left(t_c-t\right)^{-3/8}.
\end{equation}
Combining Eqs.~\eqref{eq:orphan_time_delay} and~\eqref{eq:foft}, we arrive at an expression for the frequency of an orphaned pulsar term that links to a higher-frequency merger:
\begin{equation}\label{eq:orphan_f}
    f_{\rm orphan} = \frac{5^{3/8}}{8\pi}\left(\frac{G\mathcal{M}}{c^3}\right)^{-5/8}\left[(1+\hat{k}\cdot\hat{p})\frac{D}{c}\right]^{-3/8}.
\end{equation}
This frequency shift as a function of the angular and distance dependency of the orphaned pulsar-term delay time is shown in Fig.~\ref{fig:frequency-shift-map} for an example MBHB system with total mass $2\times10^9\,\msun$ at redshift $z=1$. For distant pulsars angled away from the gravitational-wave origin, the frequency shift as seen in the orphaned pulsar term can be up to several orders of magnitude.

We can then see the multiband picture take shape. MBHB mergers in LISA or \muares will provide precise measurements of sky location and chirp mass \citep{sesana_unveiling_2021,colpi_lisa_2024}. In the case of particularly high-mass mergers in LISA, or a MBHB observed via astrometry, the chirp mass can be measured, but the signal will be very poorly localized on the sky \citep{mangiagli_massive_2022, Zhang2025}. If the distances to pulsars in the array are well-measured at the time of such an observation, the combination of pulsar distance information and precise sky localization would allow for direct computation of the pulsar-by-pulsar response of the pulsar term to the pre-merger MBHB system (assuming GR-only evolution). In this case, the per-pulsar contribution can be directly stacked across the array to achieve detection. In the absence of such precise constraints, a search could be constructed that leverages merger-derived localization information to infer all unknown pulsar distances in the array. 

Crucially, even if detection of the orphaned pulsar terms is not feasible at time of merger, such a search would be nigh-eternally archival. As the evolution of the orphaned pulsar term waveform is largely negligible on human timescales, we are free to accrue continued pulsar timing observations and any additional pulsar distance measurements made in the years or decades following the merger. Archival searches can be performed until such a time as the orphaned pulsar term signal is detected.

\section{Methods}\label{sec:methods}
To estimate the prospects for detection of the early-time MBHB contribution in PTAs, we consider the case of a well-localized MBHB, with a well-measured chirp mass, and known pulsar distances. Such assumptions are not required for detection of the signal, but to relax them would necessitate construction of a search pipeline. Given our aim in this study of establishing the concept of orphaned pulsar terms and assessing the feasibility of their detection, we deem such an undertaking outside the scope of this work. 

We compute the PTA response and sensitivity to the orphaned pulsar terms by modifying {\tt hasasia}---a realistic sensitivity curve calculator for PTAs \citep{hazboun_realistic_2019, Hazboun2019Hasasia}---to isolate the pulsar term contribution to the sensitivity calculation from the full Earth term + pulsar term case. The full formalism behind {\tt hasasia} sensitivity calculations is detailed in \citet{hazboun_realistic_2019}. The modification is straightforward, amounting to removing the Earth term in Equation (35) of \citet{hazboun_realistic_2019} to yield the expression for the response given in Eq~\eqref{eq:ophan_response} above.\footnote{The corresponding author's fork of {\tt hasasia} with these changes has been published as an archival release to Zenodo \citep{hazboun_hazboun6_2019}.}

For a given MBHB, we trace back the gravitational waveform --- assuming GR evolution as given in Eq.~\eqref{eq:orphan_f} --- to the time at which passes by a given pulsar as a function of that pulsar's distance and sky location relative to the origin of the MBHB signal. From this, we compute the frequency and amplitude of the orphaned pulsar term contribution from the MBHB in each pulsar of the array. The single-pulsar, pulsar-term-only SNR is computed for the orphaned pulsar term in each pulsar and combined in quadrature to form the full array SNR of the stacked signal. Due to the directional dependence of the orphaned pulsar term, we repeat this process for signals originating from each pixel in a Healpix \citep{gorski_healpix_2005a} grid across the sky, then compute the mean SNR across the sky. 

To estimate current-day PTA sensitivity to the orphaned pulsar term contribution, we use ``DR3Like'': a realistic simulated PTA aiming to mimic expectations for the upcoming International Pulsar Timing Array $3^{\rm rd}$ Data Release (IPTA DR3), originally produced for \citet{petrov_expectations_2025, schult_expectations_2025}. This dataset consists of 116 pulsars with 22.2 years of timing baseline. Distances of the pulsars in this simulated dataset are generally not specified; as such, we draw distances at random, assuming approximate isotropy in the Galactic plane such that $p(D)\sim D^2$ with a maximum allowed distance of 5 kpc. We neglect the contribution of the gravitational-wave background to the overall noise level of the PTA, as it is subdominant at frequencies above $\sim$10 nHz and the orphan pulsar term frequencies are generally at frequencies $f_{\rm orphan} \gtrsim 10$ nHz for systems which can evolve to merger over $\mathcal{O}$(1000s of years).

Additionally, we project forward the current-day SNRs to estimate prospects for the PTAs of the 2050s and beyond, using a simple SNR-scaling relation for the combination of single-pulsar matched filer searches:
\begin{equation}\label{eq:pta_scaling}
    \rho \propto \frac{\sqrt{N_{\rm psr}}\sqrt{T}}{\sigma_{\rm rms}},
\end{equation}
where $N_{\rm psr}$ is the number of pulsars in the array, $T$ is the total timespan of the observation, and $\sigma_{\rm rms}$ is the root-mean-square timing uncertainty.

The Square Kilometer Array (SKA) is expected to discover $\sim800$ new millisecond pulsars and perform dedicated timing on 174 known systems, augmented by newly discovered systems \citep{shannon_skao_2025}. We assume a conservative total of 200 pulsars timed by SKA, 33 of which were already present as southern-hemisphere pulsars in the DR3Like simulation, thereby yielding 167 additional pulsars. In the northern hemisphere, the Deep Synoptic Array (DSA) will time 200 pulsars \citep{hallinan_dsa2000_2019b}. We assume 83 of these are the pulsars included in the current NANOGrav PTA, yielding an additional 117 pulsars added to the array. Overall, this yields an additional 284 pulsars. We consider a projected sensitivity in 2050, i.e., for a post-LISA archival search or a contemporaneous search with a Voyage 2050 concept such as $\mu$Ares; this amounts to 25 years of additional baseline. Finally, the SNR should scale as $\rho \propto \sigma_{\rm rms}^{-1}$. Both SKA and DSA expect to increase the precision of timing measurements substantially, although the specific improvement will vary from pulsar to pulsar. Both observatories expect to achieve ns precision on a subset of pulsars, while still maintaining timing measurements of pulsars with a precision of order $\mu$s. Overall, both observatories expect a factor of at least 3-4 improvement in sensitivity over current timing arrays \citep{shannon_skao_2025,hallinan_dsa2000_2019b}. Together, these estimates --- $N_{\rm psr}=400$, $T=47.2$ years, and $\sigma_{\rm rms, 2050}=0.25\times\sigma_{\rm rms, now}$ --- comprise the post-SKA/DSA-2000 ``2050'' scenario. In the spirit of an infinitely archival search, we also speculate as to what could be achievable in the year 2100 (the ``2100'' scenario), with 100 years of timing baseline, 1000 pulsars across an international array, and with $\sigma_{\rm rms, 2100}=0.1\times\sigma_{\rm rms, now}$.

\subsection{Other Observatories}\label{sec:methods_obs}
We perform straightforward sensitivity estimates for MBHB merger SNR in higher-frequency observatories. We consider two space interferometers (LISA and \textmu Ares) and astrometric observations with \textit{Roman} and \textit{Kepler}. For a given MBHB, we compute its characteristic strain track beginning 1 yr prior to merger. We then compute the SNR with respect to the effective sensitivity:
\begin{equation}
    \rho^2 = \int_{-\infty}^{+\infty} \left[ \frac{h_c^2(f)}{fS_n(f)} \right] d\log f,
\end{equation}
where $h_c$ is the characteristic strain and $S_n$ is the effective noise power spectral density (PSD), such that
\begin{equation}
    S_n(f) = \frac{P_n(f)}{\mathcal{R}(f)},
\end{equation}
where $\mathcal{R}(f)$ is the sky- and polarization-averaged detector response. For LISA, we compute the noise PSD as given in \citet{colpi_lisa_2024} and construct $\mathcal{R}(f)$ for the X-Y-Z time-domain interferometry channels \citep{tinto_secondgeneration_2023a, tinto_timedelay_2020a}. We account for the Galactic foreground with the estimate of \citet{robson_construction_2019}. For \textmu Ares, we use the estimate of $S_n$ described in \citet{sesana_unveiling_2021}. For astrometric observations, we compute the astrometric sensitivity such that
\begin{equation}
    S_n(f) = 2 t_{\rm{cadence}} \frac{\sigma^2_{ss}}{N_{\rm{meas}} N_{\rm{stars}}},
\end{equation}
where $N_{\rm stars}$ is the total number of observed stars in the survey, $t_{\rm cadence}$ is the survey cadence, $\sigma_{ss}$ is the per-star astrometric precision, and the number of measurements $N_{\rm meas}$ is approximated here as $T_{\rm survey}/t_{\rm cadence}$ with $T_{\rm survey}$ being the survey duration \citep{Wang2021}. \textit{Kepler} archival data covers 160,000 stars over 3.5 yrs with a 30 min cadence and estimated mean astrometric precision of 0.7 mas/exposure/star \citep{Zhang2025,monet_preliminary_2010}.
The \textit{Roman} Galactic Bulge Time-Domain Survey is planned to cover $10^8$ stars over 5 yrs (1.15 yrs of dedicated time) with a cadence of 12 min and per-star astrometric precision of 1.1 mas/exposure \citep{wfirstastrometryworkinggroup_astrometry_2019, gaudi_2019_AuxiliaryScienceWFIRST, gaudi_2023_RomanGalacticExoplanet, ObservationsTimeAllocationCommittee2025}.

\subsection{MBHB Rate Simulations}\label{sec:methods_rates}
To estimate the expected rate of multiband-observable MBHB systems, we employ the \texttt{L-GalaxiesBH} semi-analytical model (SAM), run on the extended version of the Millennium merger trees presented in \cite{bonoli_constraints_2025}. The dark-matter resolution of these merger trees allows us to accurately trace the assembly of MBHs with masses $M_{\rm BH}\,{>}\,10^6 M_{\odot}$. In particular, we adopt the \textit{Super-Edd} implementation of the \texttt{L-GalaxiesBH} model described in \cite{bonoli_constraints_2025}, which has been shown to produce a nHz gravitational-wave background amplitude compatible with current PTA measurements. In brief, the SAM includes different prescriptions to track the assembly of galaxies and multiple channels for MBH formation \citep[see][]{spinoso_multiflavour_2023} and growth. The latter is regulated by distinct accretion phases spanning Eddington-limited, super-Eddington, and sub-Eddington regimes \citep[as in][]{izquierdo-villalba_connecting_2024}. \texttt{L-GalaxiesBH} self-consistently tracks the evolution of MBH spins driven by gas accretion and MBH mergers, and follows the dynamical evolution of MBH binaries from kiloparsec separations down to milliparsec scales. This is achieved through the inclusion of dedicated prescriptions accounting for dynamical friction, stellar and gaseous hardening due to interactions with single stars and circumbinary accretion discs, and gravitational wave emission. 

The outputs of \texttt{L-GalaxiesBH} consist of a complete catalog of MBH mergers occurring within the cosmological volume of the Millennium simulation, spanning from redshifts $z = 56$ to $z = 0$. These mergers are converted into a cosmological merger rate per unit redshift and chirp-mass bin according to
\begin{equation}
         \frac{dN}{dz d \mathcal{M}} = \frac{n(\mathcal{M}\pm\Delta\mathcal{M},z \pm \Delta z)}{\Delta \mathcal{M} \Delta z} \frac{4 \pi c d_L^2}{(1+z)^2},
\end{equation}
where $d_L$ is the source luminosity distance at a redshift $z$, and $n(\mathcal{M}_c \pm \Delta \mathcal{M}_c, z \pm \Delta z)$ is the comoving number density of MBHBs in a given redshift and chirp-mass bin as predicted by the SAM.

\section{Prospects}\label{sec:prospects}

We compute the stacked array SNR of the orphaned pulsar terms for MBHBs at every point in the $\{\log_{10}M-\log_{10}z\}$ grid used for the rate computation in \S\ref{sec:methods_rates}, save that we only consider systems with total mass $\geq 10^7\,\msun$ and redshift $z\leq4$. We compute the merger SNR in LISA, \textmu Ares, and \textit{Kepler}/\textit{Roman} astrometry across the same grid, assuming a conservative 1 year in-band observation time for each observatory. We consider only those systems whose merger exceeds $\mathrm{SNR}\geq10$ in at least one of the higher-frequency observatories (or $\mathrm{SNR}\geq3$ for the astrometric case). 

\begin{figure}
    \centering
    \includegraphics[width=0.75\linewidth]{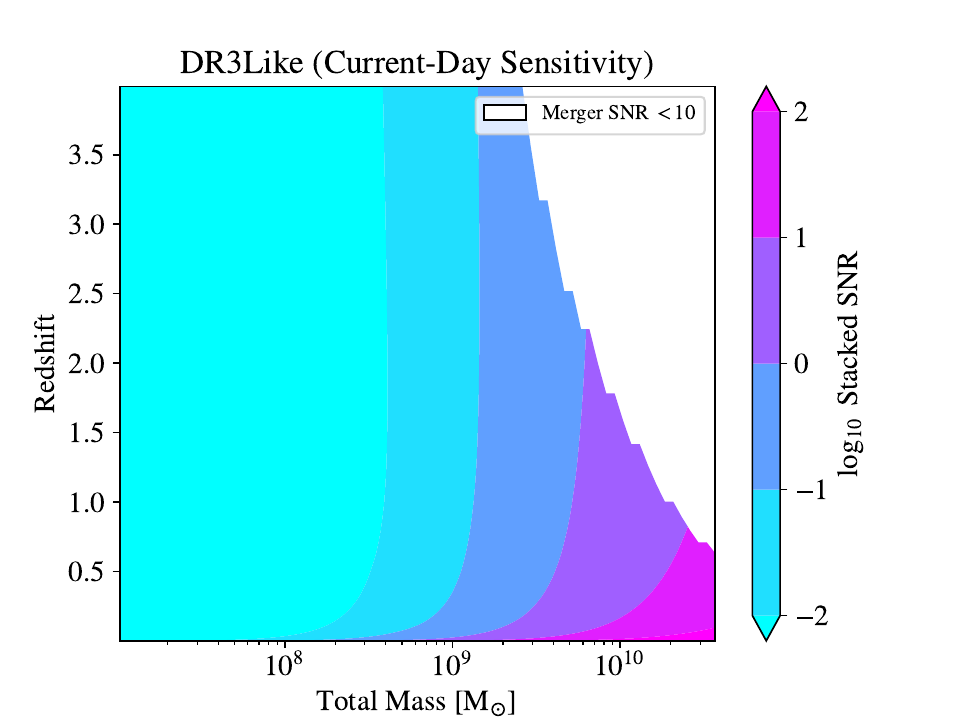}
    \caption{Sky-averaged stacked array SNR in the DR3Like simulation for orphaned pulsar term signatures of MBHBs as a function of total mass and redshift, for those signals whose mergers can be observed at higher frequencies with a SNR of at least 10. The SNRs shown are representative of current-day PTA sensitivity. }
    \label{fig:orphan_snrs_dr3like}
\end{figure}

\begin{figure*}[t!]
    \centering
    \begin{subfigure}[t]{0.5\textwidth}
        \centering
        \includegraphics[width=1\linewidth]{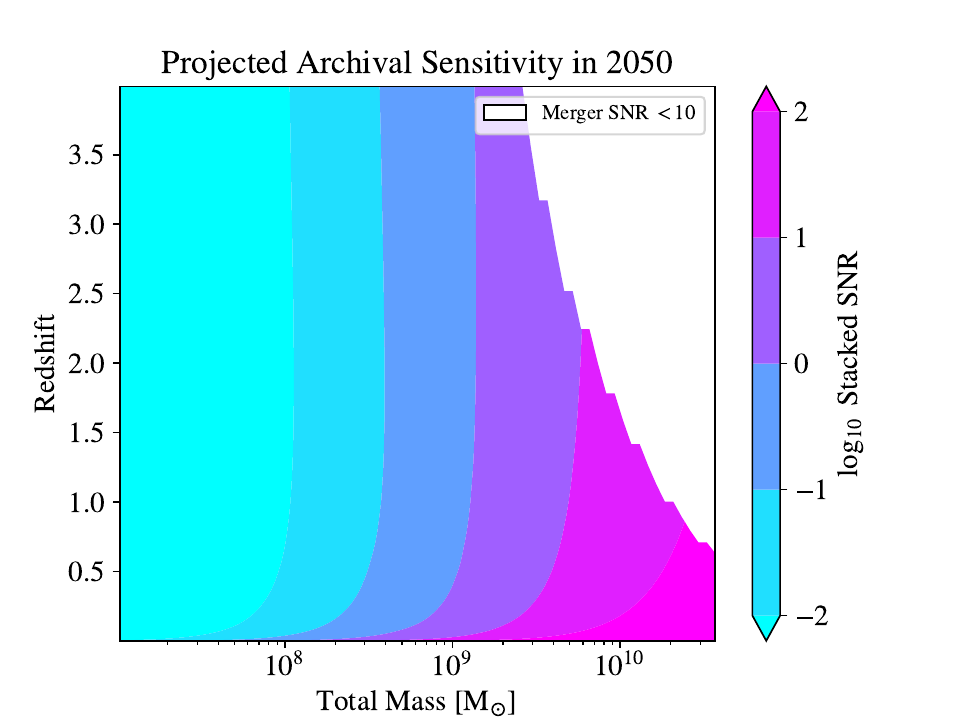}
    \end{subfigure}%
    ~ 
    \begin{subfigure}[t]{0.5\textwidth}
        \centering
        \includegraphics[width=1\linewidth]{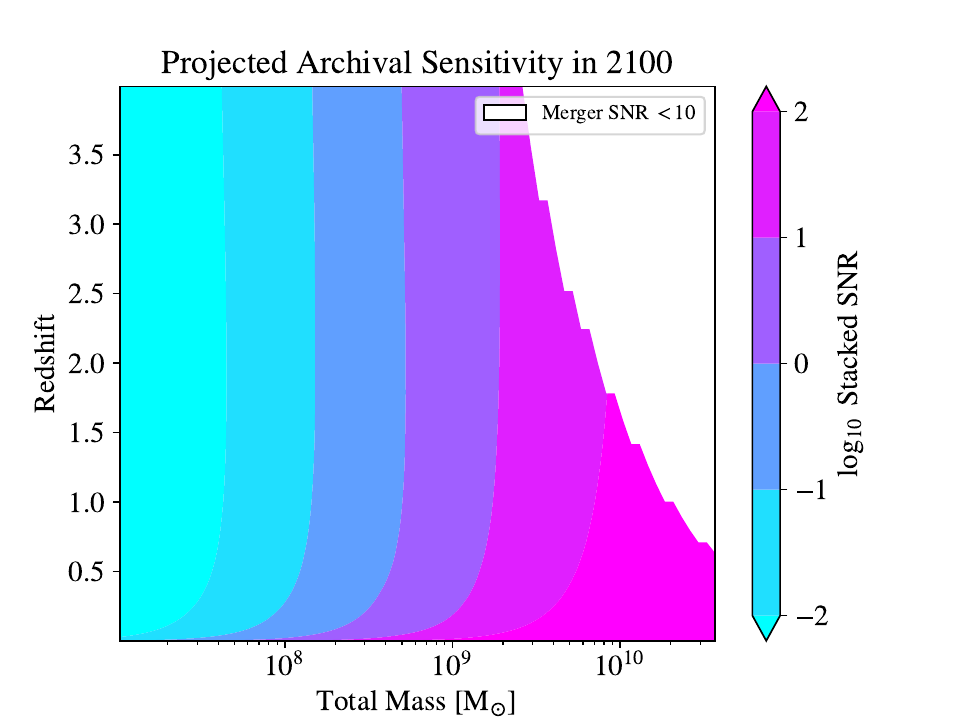}
    \end{subfigure}
    \caption{Scaled sky-averaged stacked array SNR for orphaned pulsar term signatures of MBHBs as a function of total mass and redshift, for those signals whose mergers can be observed at higher frequencies with a SNR of at least 10. The SNRs shown are a projection of current-day PTA sensitivity to what may be possible in an archival search in 2050 (left) and 2100 (right).}
    \label{fig:orphan_snrs_scaled}
\end{figure*}

The full-array orphaned pulsar term SNRs are shown in Fig.~\ref{fig:orphan_snrs_dr3like} for a simulated PTA representative of current-day sensitivity. Additionally, we estimate the stacked orphan pulsar term SNRs for the archival 2050 and 2100 scenarios discussed in \S\ref{sec:methods} by projecting forward the DR3Like SNRs via the PTA scaling relation given in Eq.~\eqref{eq:pta_scaling}. The resulting scaled SNRs as a function of $M$ and $z$ for these scenarios are shown in Fig.~\ref{fig:orphan_snrs_scaled}. It is immediately apparent that the orphaned pulsar term mechanism is most relevant for high-mass systems, i.e. $M\gtrsim10^9\,\msun$. Both LISA and \muares provide opportunities for multiband detections of such systems, with \muares providing significantly further reach in terms of redshift ($z\lesssim3$ for \muares as opposed to $z\lesssim0.5$ for LISA). Astrometric detection is constrained to the extreme high-mass end of the distribution ($M\gtrsim5\times10^9\,\msun$, up to $z<0.05$ at $M=10^{10}\,\msun$), but has the unique property that an astrometrically-detectable merger is guaranteed to be accompanied by high-SNR orphaned pulsar terms.

With the MBHB merger rate simulations described in \S\ref{sec:methods_rates}, we compute the expected detection rate for the orphaned pulsar term signal, considering again the three scenarios outlined above. These rates can be found in Table~\ref{tab:rates}. These rates rely on our assumption of equal-mass, circular systems with solely gravitational-wave-driven evolution. Relaxing these assumptions will likely affect the rates quoted here; this is discussed further in \S\ref{sec:discussion}.

The occurrence rate of circular multiband systems via the orphan pulsar term mechanism is overall low, mostly due to the rarity of high-mass MBHB mergers. While systems exist which could in principle be observed with current-day PTA sensitivity, we estimate a negligible probability of detecting such a signal at present. However, recall the effectively infinitely archival nature of this multiband signal class. The orphaned pulsar terms of any near-future MBHB merger are already present in the PTA datastream; to improve these chances one must merely wait for a detected merger, accruing additional PTA baseline --- and therefore SNR --- in the meantime. By 2050, LISA will have flown, the \textit{Roman} survey will be complete, and both SKA and DSA will have provided extraordinarily precise pulsar timing data, bringing the probability of multiband detection with (PTA) SNR$\ge3$ up to $0.5-1$\%. With further accrual of pulsar timing data and the advent of \textmu Ares, this increases to $4-8$\% by 2100. It is worth noting that these prospects may be enhanced in the case of eccentric binaries; see \S\ref{sec:astro_eccentricity} for further discussion of this point. 

Finally, note that the quoted detection probabilities are for the \textit{multiband} case specifically; the orphaned pulsar terms will be present regardless of whether we happen to observe the merger itself. Indeed, due to the long timescales of light travel time in the PTA array, SNR$\ge3$ orphaned pulsar terms from a MBHB merger --- a 1 in 45,000 yr event --- have a $\sim50$\% chance to be present across at least some of the pulsars in the array at current-day sensitivity. For the 2050 scenario, we expect $16^{+7}_{-6}$ orphaned systems with SNR$\ge3$ to be present within the array. However, finding such systems without knowledge of their chirp mass or sky position --- and therefore frequency evolution --- via a multiband counterpart will be highly challenging. In fact, care will need to be taken so as not to claim a false multiband detection by mistaking other systems' orphaned pulsar terms for the PTA counterpart to a promising high-frequency merger. That being said, the specific frequency response of the orphaned pulsar term mechanism for a given MBHB mass and direction of origin can be expected to alleviate this somewhat --- at least up to uncertainty on pulsar distances.

\begingroup
\setlength{\tabcolsep}{10pt} 
\renewcommand{\arraystretch}{1.5} 
\begin{table}
    \centering
    \begin{tabular}{c|c|c|c}
        Scenario & Rate & Inverse Rate & $p({\rm multiband\ detection})$ \\
        \hline
        DR3Like & 2.2$\times10^{-5}$/yr & 1 per 45,000 yr & 0\%\\
        2050 & 4.9$\times10^{-4}$/yr & 1 per 2000 yr & 0.6 -- 1\%\\
        2100 & 3.6$\times10^{-3}$/yr & 1 per 280 yr & 4 -- 8\% \\
    \end{tabular}
    \caption{Multiband detection rates and associated probability for an array-stacked orphan pulsar term PTA SNR $\ge3$. We estimate the probability of detection by considering a Poisson process with effective rate equal to the product of the per annum rate and the total years of observing time for the higher-frequency observatories. We perform this calculation for both a conservative and an optimistic set of assumptions. For the conservative (optimistic) case, the estimated probability of a multiband detection assumes 4 (10) years of LISA, 4 (10) years of muAres observations for the 2050 and 2100 cases, and 4.65 years of astrometric baseline between \textit{Kepler} and \textit{Roman}. Note that these quotes are conditioned on joint multiband detection. It is quite likely that mergers will occur in the Universe which leave orphaned pulsar terms in our array; it is substantially less likely that we will happen to observe such an event during the limited time in which higher-frequency observatories are operating.}
    \label{tab:rates}
\end{table}
\endgroup

\section{Astrophysical Implications}\label{sec:astro}
While likely rare, multiband observations of MBHBs via the orphaned pulsar term mechanism have the potential to provide unique and novel insight into a breadth of astrophysics, including current-day constraints on the presence of individual astrometrically-observable MBHBs, the precise measurement of pulsar distances, insights into MBHB environments and their evolution over time, and the prevalence and origins of MBHB eccentricity. 

\subsection{Implications for Astrometric Searches}
An interesting facet of these results is that \textit{if} a MBHB merger is detected in astrometric data, it \textit{will} be observable as a high-SNR orphaned pulsar-term contribution in PTAs. Even at present-day sensitivity, the orphaned pulsar terms associated with an astrometric merger would have a significant collective SNR ($\gtrsim10-100$ across the array). This presents immediate utility for multiband science with the signal class discussed in this work: as an independent verifier --- or falsifier --- of a candidate astrometric gravitational wave detection.

First, we note that our treatment of the astrometric case is quite conservative; by standardizing our treatment of all higher-frequency observatories to assert a) a merging system, which b) is in-band for 1 yr prior to merger, we likely underestimate the multiband prospects in connection with astrometric observations which --- much like PTAs --- are a long-term experiment. Even so, astrometry currently has much lower sensitivity than the other high-frequency observatories that we consider. If a high-mass MBHB were to be present in astrometric data, it would likely only be detectable with low confidence. Such a low-confidence signal candidate could be produced by systematics or other noise processes, and may be difficult to verify. However, even a tentative (SNR $\sim 3$) MBHB candidate in astrometric data could be immediately confirmed or rejected by a search for the associated orphaned pulsar terms. This mechanism therefore provides an independent check on any future astrometric signal candidate.

Conversely, if an open search were to be performed for orphaned pulsar terms in current-day PTAs, it would set independent constraints on the presence of individually-observable MBHB mergers in astrometric data and therefore the rate of high-mass mergers. Such an open search would  be quite challenging absent constraints from a higher-frequency merger, and we leave consideration of a non-multiband orphan pulsar term search to future work.

\subsection{Pulsar Distances}
While in this work we assume good knowledge of pulsar distances for the sake of convenience, this is generally not be the case in practice \citep[see e.g.,][]{manchester_australia_2005}. Instead, a realistic orphaned pulsar term search would also infer the pulsar distances of every pulsar in the array. In the case of a circular MBHB with purely gravitational-wave driven evolution, these inferred distances could be extraordinarily precise --- indeed, this was the focus of \citet{pitkin_there_2008}. Beyond the immediate benefits of this information for pulsar timing itself, such measurements would also be a boon to a wide variety of pulsar-derived science, including studies of the interstellar medium and improved characterization of binary pulsar periods, the latter of which would enable improved precision in tests of general relativity and binary pulsar mass measurements.

\subsection{Environmental Effects}
A major question which we hope to answer with LISA is that of the presence and nature of accretion disks around MBHBs \citep{amaro-seoane_astrophysics_2023e}. Gas in the accretion disk mainly interacts with the binary via gravitational torques; depending on the nature of the disk (i.e., its temperature, surface density, and scale height), this torque can either inject or remove angular momentum from the binary \citep{haiman_population_2009}. This is realized in the MBHB waveform as an additive term to the frequency evolution of the binary, i.e. $\dot{f} = \dot{f}_{\rm GW} + \dot{f}_{\rm gas}$. The former term describes the vacuum evolution of the system, whereas the latter term encodes the effects of gas torques and can be positive or negative. These effects are difficult to characterize when considering only a MBHB merger, as they will induce only a small perturbation to the merger waveform --- if they leave any imprint at all \citep{garg_measuring_2024}. 

The orphaned pulsar term mechanism provides a natural means by which to characterize these processes, as pulsars across the array view the system at discretely-sampled times over tens of thousands of years. The transition between the gas-dominated to the GR-dominated regime likely occurs either while or just after the MBHB signal is present in the orphaned pulsar terms. While this presents a search challenge, it indicates that the MBHB environment can have a strong influence on orphaned pulsar term contributions. Crucially, tracing the MBHB's frequency evolution across pulsars in the array \citep{2012PhRvL.109h1104M} should provide a conclusive test of the presence or absence of a gaseous accretion disk, and allow for constraints on $\dot{f}_{\rm gas}$ via deviations of the observed frequency evolution from $\dot{f}_{\rm GW}$ as inferred from the merger. Outside of multimessenger observations of a confirmed electromagnetic counterpart to the gravitational-wave detection of a MBHB merger, the orphaned pulsar term mechanism may provide our only opportunity to directly observe the presence and impact of a gaseous disc around a merging MBHB. It is worth noting, however, that such measurements will be degenerate with our uncertainty as to the pulsar distances; absent additional, external distance constraints it may be challenging to achieve meaningful constraints.

\subsection{Eccentric Binaries}\label{sec:astro_eccentricity}
Several mechanisms have been proposed to induce eccentricity in MBHBs, and it may be that eccentric MBHBs are in fact common \citep{2011MNRAS.415.3033R}. These can range from core scouring and prior mergers in the host galaxy \citep{2025A&A...694A.282T}, to interactions of the MBHB and a circumbinary accretion disk prior to the gravitational wave dominated regime \citep{2011MNRAS.415.3033R}, to eccentricity of the original galaxy merger \citep{2024MNRAS.532..295F}. By characterizing these systems' eccentricity, we can begin to understand more facets of their formation and evolution. Unfortunately, detecting the presence of eccentricity in observatories like LISA and \muares is quite challenging; while some residual eccentricity may be detectable \citep{garg_minimum_2024}, in many cases the binary will have circularized via gravitational wave emission prior to merger. PTA datasets, however, have been noted as a possible means to explore the eccentricity of the overall MBHB population \citep[e.g.,][]{2025A&A...703A..86F,2025arXiv250614882M,2025PhRvD.111b3047F}.

The archival multiband signal class discussed in this work presents a unique opportunity to study the eccentricity of MBHBs, even in the case of nondetection. Due to the coupling between the eccentricity $e$ and the frequency evolution of the system, non-zero eccentricity will enhance the frequency shift between the merger and the orphaned pulsar term, such that for the $n^{\rm th}$ harmonic,
\begin{equation}\label{eq:fdot_of_e}
    \dot{f}_n = \frac{48n(G\mathcal{M})^{5/3}}{5\pi c^5}(\pi f)^{11/3} F(e),
\end{equation}
where $F(e)$ is defined per \citet{peters_gravitational_1963} as
\begin{equation}
    F(e) = \frac{1+(73/24)e^2 + (37/96)e^4}{(1-e)^{7/2}}.
\end{equation}
Note that the eccentricity also evolves through gravitational wave emission and as such solutions for the differential equation in Eq.~\eqref{eq:fdot_of_e} must be approximated via numerical integration. Moreover, the eccentric MBHB's gravitational-wave power and evolution are split across its harmonics, further complicating calculation of its orphaned pulsar term contribution. Due to the associated computational and analytic challenge, we leave a detailed treatment of orphaned pulsar terms for eccentric MBHBs to future work. 

\begin{figure}
    \centering
    \includegraphics[width=0.9\linewidth]{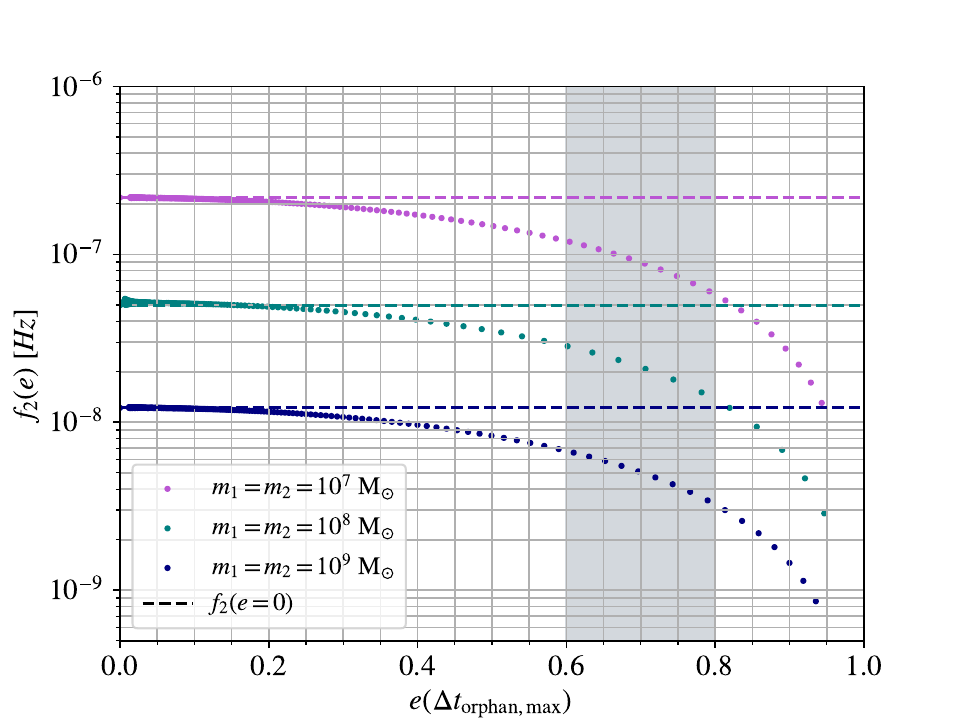}
    \caption{Gravitational-wave frequency of the $n=2$ harmonic in the pulsar with the greatest $\Delta t_{\rm orphan}$, as a function of eccentricity (as observed in the same pulsar), for three different (source frame) MBHB masses. The dotted points are the result of numerical integration of the eccentric evolution equations, normalized to the same merger time $t_c$. The dashed lines provide the (monochromatic) gravitational-wave frequency for the circular case as a reference. The grey-shaded region encompasses the range of limiting eccentricities for MBHB systems with circumbinary accretion disks \citep{2009MNRAS.394.2255S}. For eccentricities in this regime, the orphaned pulsar term contribution can be pushed to substantially lower frequencies --- thereby increasing the PTA sensitivity to these terms. At the upper end of the range of limiting eccentricities, the eccentricity enhancement to the orphan pulsar term mechanism is at most equivalent to an order of magnitude increase in mass.}
    \label{fig:fshift-ecc-mass}
\end{figure}

Here, however, it is useful to establish a quantitative expectation of the eccentricity-enhanced frequency evolution over timescales accessible to pulsars in our array. To that end, we consider the (source-frame) effects of eccentricity on equal-mass MBHB systems with total masses $M\in[10^7,\ 10^8,\ 10^9]\,\msun$ using the {\tt LEGWORK} package \citep{wagg_legwork_2021} for numerical integration and attendant calculations. In Fig.~\ref{fig:fshift-ecc-mass}, we show the enhancement relative to the circular case of the frequency of the $n=2$ harmonic in the orphaned pulsar term with the largest $\Delta t_{\rm orphan}$ --- i.e., in the pulsar which observes the MBHB system at the earliest time in its evolution --- as a function of the eccentricity observed in that pulsar. It is important to note that this only sets a lower bound on the frequency of the orphaned pulsar term, as for highly-eccentric systems much of the signal power is emitted across higher harmonics. Additionally, the signal morphology of such eccentric systems becomes deeply complicated; detection of orphaned pulsar terms from eccentric systems --- especially absent good constraints from the likely-circularized merger --- will be quite challenging.

Generally speaking, $e \gtrsim 0.5$ is required to significantly lower the frequency of the orphaned pulsar term relative to the circular case. For highly-eccentric MBHBs with $e\gtrsim0.9$, the enhancement can be nearly an order of magnitude in frequency. While lower eccentricities do not provide the benefit of increased frequency-evolution, it is crucial to note that in such cases the eccentricity of the system would be observable by PTAs but will have circularized entirely by the time of the merger. Multiband detection of such a system via orphaned pulsar terms would likely be our only opportunity to study its eccentricity. Finally, the effect of eccentricity on the orphaned pulsar term mechanism has the interesting consequence that the non-detection of orphaned pulsar terms in connection with an observed MBHB merger may in some cases place upper limits on the eccentricity of the system, furthering the multiband utility of performing these kinds of searches.

Notably, \citet{2011MNRAS.415.3033R} conclude that in the presence of a circumbinary accretion disk, MBHBs tend to evolve towards a limiting eccentricity $e_{\rm crit}\in[0.6,0.8]$ prior to entering the regime where gravitational wave emission dominates (note that other works find a limiting eccentricity around 0.4-0.5 \citep{2021ApJ...914L..21D,2023MNRAS.522.2707S}, although employing 2D, isothermal simulations). As seen in Fig.~\ref{fig:fshift-ecc-mass}, this interval corresponds to significantly enhanced frequency shifts in the orphaned pulsar terms for a given MBHB mass, potentially improving the PTA sensitivity to lower-mass systems. If the majority of MBHBs do indeed possess initial eccentricities close to $e_{\rm crit}$, it would carry significant implications for the prospects of low-frequency multiband observations. Access to lower-frequency orphaned pulsar term contributions from less-massive systems could enhance the rates discussed in \S\ref{sec:prospects}, simply through improved PTA sensitivity at the relevant frequencies. Indeed, much of the scaling between scenarios in Table~\ref{tab:rates} is due to increased pulsar-term SNR of lower-mass MBHBs, which occur at substantially higher rates than high-mass systems; eccentricity enhancement of the MBHB frequency evolution will give us earlier and more plentiful opportunities for multiband detection compared to other regions of the parameter space. That being said, eccentricity remains a double-edged sword; it is not clear whether the added opportunities it presents can make up for the substantial complexity it would induce for an orphaned pulsar term search.

\section{Discussion}\label{sec:discussion}
We present a detailed consideration of the prospects and implications of low-frequency multiband observations of MBHBs via the orphaned pulsar term mechanism. This novel approach allows for connection of MBHB mergers detected with astrometric or space-interferometric observatories to their lingering contributions in the PTA band. These orphaned pulsar terms act as time-delayed probes of the signal, tracing it not as it existed at merger, but rather thousands of years beforehand. Crucially, while the merger itself is a transient event, the orphaned pulsar terms evolve on significantly longer timescales than PTA observations; this allows for archival searches and detection long after the merger itself is observed. This kind of archival, multiband signal has the potential to trace the effects of MBHB environments on their evolution over thousands of years, provide otherwise unattainable insight into their eccentricities, and enable additional pulsar science via precise pulsar distance estimates. Even in the event of non-detection, the mechanism can still provide astrophysically-meaningful constraints on astrometric observations, the rate of high-mass MBHB mergers, and MBHB eccentricity. Finally, due to the fact that an astrometrically-detected MBHB is always accompanied by a high-SNR orphaned pulsar term contribution, the multiband search proposed in this work can provide independent verification --- or falsification --- of a candidate astrometric gravitational-wave detection.

For equal mass, circular MBHBs, the prospects for detection of such a signal are overall low in the immediate future: $0.5-1$\% by 2050. These rates are largely driven by the relative dearth of massive $(M \gtrsim10^9\,\msun)$ MBHB mergers in our MBHB rate simulation. The archival nature of the search allows for some accrued probability of detection following the merger; this figure improves to $4-8$\% by 2100. In contrast to transient phenomena, the scaling of this archival search with time is primarily driven by expanded access to lower-mass systems via longer integration times. It is important to note that a low probability of multiband detection does not indicate absence of orphaned pulsar terms in PTAs. In fact, the longevity of this mechanism will likely result in a not insignificant number of near-threshold orphaned MBHBs strewn across the array. While challenging to detect individually without multiband information, these systems may give rise to a likely non-Gaussian astrophysical pulsar-term-only noise which eschews the usual correlation patterns we expect from the nHz gravitational-wave background. This angle presents a compelling direction for future work. 

Additionally, the orphaned pulsar term mechanism may be enhanced in the case of eccentric systems. This is due to the increased gravitational-wave emission (and therefore extraction of orbital energy) in the eccentric case relative to a circular binary, resulting in faster frequency evolution over time and an increased $\Delta f_{\rm orphan}$. While a detailed treatment of the orphaned pulsar term search for eccentric binaries is beyond the scope of this initial work, we provide estimates of the eccentricity enhancement to the orphaned pulsar term frequency. We show that for $e\in[0.6,0.8]$ the enhancement is at most equivalent to an order of magnitude in mass, although the power of such a system will be distributed across higher-frequency modes. As PTA sensitivity increases rapidly with reduced frequency, multiband detection prospects may be improved for eccentric binaries (depending on the level of impact of the necessarily-increased search complexity in such a case). Conversely, non-detection of the orphaned pulsar term may allow PTAs to set upper limits on the initial eccentricity of MBHBs observed by LISA or \textmu Ares. Smaller eccentricities do not provide a significant enhancement to the orphaned pulsar term mechanism, but may comprise the only means by which we can investigate the eccentricity of a MBHB that has circularized by the time it reaches the LISA band. Additionally, intermediate mass-ratio inspirals (IMRIs) consisting of a smaller MBH $M\sim10^4-10^6\,\msun$ and a supermassive black hole will also give rise to orphaned pulsar term contributions and may be relatively more common \citep{sedda_merging_2021}.

Finally, \citet{blas_bridging_2022, foster_discovering_2025} have proposed the concept of detecting gravitational waves via binary pulsar resonances. This kind of array \textit{only} consists of pulsar terms, and as such has access to the orphaned pulsar term mechanism. A promising aspect of a binary resonance array is that its sensitivity lies in the \textmu Hz band. It would therefore be sensitive to orphaned pulsar terms from significantly lower-mass MBHBs, resulting in increased rates and better prospects for LISA-detected mergers. These aspects and others will be explored in future work.

\backmatter

\bmhead{Supplementary information}
All code and data products will be made publicly available upon publication via an archival
release on Zenodo.

\bmhead{Acknowledgements}
The authors would like to thank Andrea Derdzinski and Matthew Miles for several helpful conversations. SRT thanks Yuri Levin for helpful discussions during the 32nd Texas Symposium on Relativistic Astrophysics. AWC acknowledges support by NSF NRT-2125764. SRT acknowledges support from NSF AST2307719, an NSF CAREER PHY-2146016, and a Vanderbilt University Chancellor's Faculty Fellowship. AWC and SRT acknowledge support from the Vanderbilt Office of the Vice Provost for Research \& Innovation's Fall 2024 Seeding Success Grant: ``Bridging The Gulf Between Solar-system-sized and Galaxy-sized Gravitational Wave Detectors''. AWC and SRT are members of the NANOGrav collaboration, which receives support from NSF Physics Frontiers Center award numbers 1430284 and 2020265. This work was conducted in part using the resources of the Advanced Computing Center for Research and Education (ACCRE) at Vanderbilt University, Nashville, TN.  S.B. acknowledges support from the Spanish Ministerio de Ciencia e Innovaci\'on through project PID2024-159201NB-C21. DIV tanals La Caixa foundation, as the project that gave rise to these results received the support of a fellowship from ``la Caixa'' Foundation (ID 100010434). While finalizing this manuscript, we became aware of concurrent efforts exploring related concepts (Zheng et al., in prep.; Quelquejay-Leclere et al., in-prep.) and agreed to coordinate release of these otherwise independent manuscripts.

\bibliography{NG_Multiband_Echoes,Multiband_Addtl_Refs,Code_Packages}

\end{document}